\documentclass[12pt]{article}
\usepackage{amsmath}
\usepackage{amssymb}

\ifx\pdftexversion\undefined
  \usepackage[dvips]{graphicx}
\else
  \usepackage[pdftex]{graphicx}
\fi

\advance\oddsidemargin by -0.5in 
\advance\textwidth by 1in
\begin{document}

%-------------------------------------------------------------------------------------------
%-------------------------------------------------------------------------------------------
\begin{titlepage}
%-------------------------------------------------------------------------------------------
%-------------------------------------------------------------------------------------------
 \date{}
% \title{Report-2 (April, 2004). Quantum Tunneling.}
\title{Magnetic relaxation of superconducting quantum dot: two-dimensional false vacuum decay}
 \author{D. R. Gulevich and F. V. Kusmartsev \\
 \\
 {\it Department of Physics, Loughborough University,} \\
 {\it Loughborough, Leicestershire LE11 3TU, United Kingdom} \\}
 \maketitle
 \begin{abstract}
 Quantum tunneling of vortices has been found to be an important novel phenomena for description of low temperature creep in high temperature superconductors (HTSCs). We speculate that quantum tunneling may be also exhibited in mesoscopic superconductors due to vortices trapped by the Bean-Livingston barrier. The London approximation and method of images is used to estimate the shape of the potential well in superconducting HTSC quantum dot. To calculate the escape rate we use the instanton technique. We model the vortex by a quantum particle tunneling from a two-dimensional ground state under magnetic field applied in the transverse direction. The resulting decay rates obtained by the instanton approach and conventional WKB are compared revealing complete coincidence with each other.

 \end{abstract}
 \setcounter{page}{0}
\end{titlepage}

\newpage

%\section{Quantum escape from superconducting quantum dot}

%-------------------------------------------------------------------------------------------
%-------------------------------------------------------------------------------------------
\section{Introduction}
%-------------------------------------------------------------------------------------------
%-------------------------------------------------------------------------------------------

\indent\par 
  Quantum tunneling attracts much interest because of its importance for such physical systems as helium surface structures, quantum dots, superconductors, that are among the most promising for realization of large scale quantum computers. Being one of the most important manifestation of quantum mechanics, tunneling process occurs in many phenomena in physics, chemistry and biology.

  Large progress has been made in description of tunneling decay when the instanton technique was applied. Being a nonperturbative method, it plays a fundamental role in description of such type of processes ~\cite{Coleman}. It became a powerful tool in physics and has got many applications. Some of the advantages of the instanton technique (IT) could be: (i) Stronger than standard WKB. Generally, it is hard to tell whether WKB result is accurate, whereas the IT is controlled by well defined expansion parameters. (ii) No connection formulas, in some cases it is more accurate~\cite{Gildener}. (iii) Non-perturbative approach. (iv) Instantons, as elementary excitations are topological objects: configurations with different number of instantons are topologically distinct. (v) Crucial importance for higher dimensional field theories.

The concept of quantum tunneling of vortices in superconductors (e.g. Refs in the review~\cite{Review} about magnetic relaxation in HTSCs) first appeared when measurements of magnetic relaxation at ultralow temperatures have been made~\cite{mag relaxation}. The experiments have shown that the relaxation rate does not disappear at zero temperature. This phenomenon was attributed to the quantum tunneling. Until now it is not well understood yet. Neither vortex mass nor Hall coefficient are known exactly. We speculate that the vortex quantum tunneling may be also exhibited in mesoscopic superconductors due to trapping potential formed by the surface barrier. This would result in higher vortex expulsion magnetic fields than expected from pure thermodynamical considerations.

Due to the Magnus force affecting the vortex dynamics in superconductor, the situation is analogous to that of a particle in magnetic field. The problem of a charged particle tunneling in presence of magnetic field has itself both theoretical and practical interest. As the tunneling is strongly affected by magnetic field, applied in transverse direction, this could be efficiently used to control qubits in possible quantum computer realizations of the future. In absence of magnetic field the decay rate is related to the imaginary part of the free energy as $\Gamma=\frac{2}{\hbar}\mathit{Im} F$ ~\cite{Langer}. The escape rate can be found making semiclassical approximations in the Euclidean path integral. The term corresponding to the ground state gives the greatest contribution to the propagator transformed to imaginary times in the limit of large time interval. It makes possible to determine the imaginary part of the ground state energy. The same considerations must be valid when the magnetic field is applied. However, because of the broken time-reversal symmetry, a complex action appears under the path integral, when transformed to Euclidean space. Furthermore the imaginary time trajectories which extremize the action become complex and the operator corresponding to the second variation of the action is a non-hermitian one and as a result possesses complex eigenvalues~\cite{Dykman}. In this case one could make analytic continuation of the path integral to a complex coordinate space or change the time contour in the complex plane. Nevertheless we show that it is possible to make analytic continuation in cyclotron frequency to transform the action to a real one. This makes the task much more transparent because of analogy with common classical mechanics where everyone has got a good physical intuition. The coincidence with the result obtained by usual WKB technique~\cite{LL,IQFT} could also serve as a proof of validity of the method. Of course it makes sense to make such trick only if we consider the task analytically, rather than numerically. Otherwise we could not "return back" to real cyclotron frequencies after having got a numerical result with the real action.
   
Normally the polar coordinates are used to study the systems that possess rotational invariance. However, it becomes hard to work with path integrals in curved coordinates, because of additional terms appearing in the action~\cite{polar, Kleinert}. Usually one always begins with time-sliced path integral in cartesian coordinates before transformed to the curved ones, since change of variables in path integrals is not a direct procedure. Everywhere in this paper we work with path integrals written in orthogonal coordinates.

%-------------------------------------------------------------------------------------------
%-------------------------------------------------------------------------------------------
\section{Abrikosov vortex in superconducting quantum dot}
%-------------------------------------------------------------------------------------------
%-------------------------------------------------------------------------------------------

%-------------------------------------------------------------------------------------------
%-------------------------------------------------------------------------------------------
\subsection{Surface barriers}
%-------------------------------------------------------------------------------------------
%-------------------------------------------------------------------------------------------
\indent\par  The barrier near the surface of type-II superconductors was first studied by Bean and Livingstone~\cite{BL}. It arises from competition of two forces: attraction two the image antivortex near the border and interaction with the Meissner current. Surface roughness is believed to suppress the barrier. However, one can notice that the influence of the surface irregularities is much less pronounced for leaving the barrier than for entry~\cite{BL}. 
\par Another possible source affecting flux dynamics in HTSCs in transverse magnetic field is the geometrical barrier~\cite{GMB}. However, it is not exhibited in superconductors of disk form. Moreover, it is expected to dominate magnetic behavior of HTSCs of flat non-elliptic form at elevated temperatures.

%-------------------------------------------------------------------------------------------
%-------------------------------------------------------------------------------------------
\subsection{Dissipation}
%-------------------------------------------------------------------------------------------
%-------------------------------------------------------------------------------------------
\indent\par There are two main forces affecting vortex dynamics in superconductors: Magnus (Hall) force and dissipation. Feigel'man et al.~\cite{Feigel'man} proposed that the Magnus force is dominant in clean superconductors, while other authors~\cite{van Dalen} argued that the vortex tunneling may occur in an intermediate regime. As long as we consider the HTSCs, the dissipative term must not be crucial because of small coherence lengths (as well as vortex cores). Indeed, the evidence for a low dissipation regime in cuprate superconductors has been presented by~\cite{evidence}.

%-------------------------------------------------------------------------------------------
%-------------------------------------------------------------------------------------------
\subsection{Vortex mass}
%-------------------------------------------------------------------------------------------
%-------------------------------------------------------------------------------------------
\indent\par The same authors~\cite{evidence} argued that the Magnus force is also smaller than standard estimates~\cite{smaller Magnus}. Thus, the mass of vortex can be relevant to the low-temperature physics of clean HTSCs in superclean limit~\cite{Chudnovsky} and should be taken into account in our model. In 1965 two contributions to the vortex mass were calculated by Suhl~\cite{Suhl}: due to the kinetic energy of the vortex core and due to electromagnetic energy. Recently, Chudnovsky and Kuklov~\cite{Chudnovsky} have shown that transversal displacements of the crystal lattice can give a significant contribution to the vortex mass. This contribution must be crucial in metals with high concentration of superconducting electrons. In our case of small coherence lenght $\xi$, the most important contribution to the mass arises from the quantization of the electron states inside the vortex core (the same paper~\cite{Chudnovsky}). It has been shown to exceed the core mass by the factor $(\epsilon_F/\Delta)^2$ (Refs [7-9] in~\cite{Chudnovsky}).

\bigskip
%-------------------------------------------------------------------------------------------
%-------------------------------------------------------------------------------------------
\subsection{Physical model: summary of basic points}
%-------------------------------------------------------------------------------------------
%-------------------------------------------------------------------------------------------
\indent\par 1. Mesoscopic HTSC disk or quantum dot at low temperature.
\par 2. Tunneling of a single point vortex trapped by the Bean-Livingston barrier.
\par 3. No surface roughness: irregularities on the edges are less important for leaving than for entry.
\par 4. No geometrical barrier.
\par 5. No dissipation: "superclean" limit. 
\par 6. No bulk pinning.
\par 7. Vortex mass is relevant.
\par 8. Magnus force is relevant.
\bigskip
\par {\it Typical parameters}:
\par Disk diameter $\sim 10-100\; nm$
\par Thickness $\sim 1\; nm$
\par Coherence length $\xi\sim 1\; nm$
\par Bulk penetration depth $\lambda\sim 100\; nm$

%-------------------------------------------------------------------------------------------
%-------------------------------------------------------------------------------------------
\section{The model}
%-------------------------------------------------------------------------------------------
%-------------------------------------------------------------------------------------------
\indent\par The system of a vortex trapped in HTSC quantum dot is analogous to that of a charged particle in 2D potential well with magnetic field applied in transverse direction. For simplicity we assume $\hbar=1$ and $m=1$ and consider the limit of large time $T$ (i.e. small decay rates), that is usual for the instanton technique. We take the potential in the form of rotationally symmetric inverted double well. The case of a particle trapped inside one-dimensional inverted double well is studied in details by~\cite{Schulman}. See also~\cite{IQFT} and~\cite{Coleman} for 1D tunneling problems with potential of another shapes. 

The Lagrangian of our 2D model in the Poincare gauge (which coincides with the Coulomb gauge in this case) $\vec{A}=(-\frac{y B}{2},\frac{x B}{2},0)$:
\begin{equation}
L=\frac{\dot{x}^2+\dot{y}^2}{2}-\frac{\omega_c}{2}(\dot x y - \dot y x) - U(r)
\label{2Dmodel}
\end{equation}
$$
U(r)=\frac{\omega^2}{2}r^2-\alpha r^4=\frac{\omega^2}{2}(x^2+y^2)-\alpha (x^2+y^2)^2
$$
Here $\omega$ denotes the frequency at the bottom of the parabolic potential well, while $\omega_c=\frac{eB}{mc}=\frac{eB}{c}$ is the cyclotron frequency.

The survival amplitude at the bottom of the well expressed in terms of the Feynman path integral:
\begin{equation}
G(\vec{0},T;\vec{0},-T)=<\vec{0}|e^{-i 2 H T} |\vec{0}> =\int \mathcal{D} \vec{r}(t) e^{i\int_{-T}^{T}L dt}
\label{G}
\end{equation}
implying the coordinates of the center by $\vec{0}$.
Transforming to imaginary times $t\rightarrow -i\tau$:
$$
<\vec{0}|e^{- 2 H T} |\vec{0}> =\int \mathcal{D} \vec{r}(\tau) e^{-\int_{-T}^{T}L d\tau}
$$
the Lagrangian transforms to
$$
L=\frac{\dot{x}^2+\dot{y}^2}{2}+i \frac{\omega_c}{2}(\dot x y - \dot y x) + U(r)
$$
As it was mentioned early the action acquires a complex part after transforming to imaginary time, in contrast with the case of zero magnetic field. 

\begin{figure}[htb!]
\begin{center}
\includegraphics{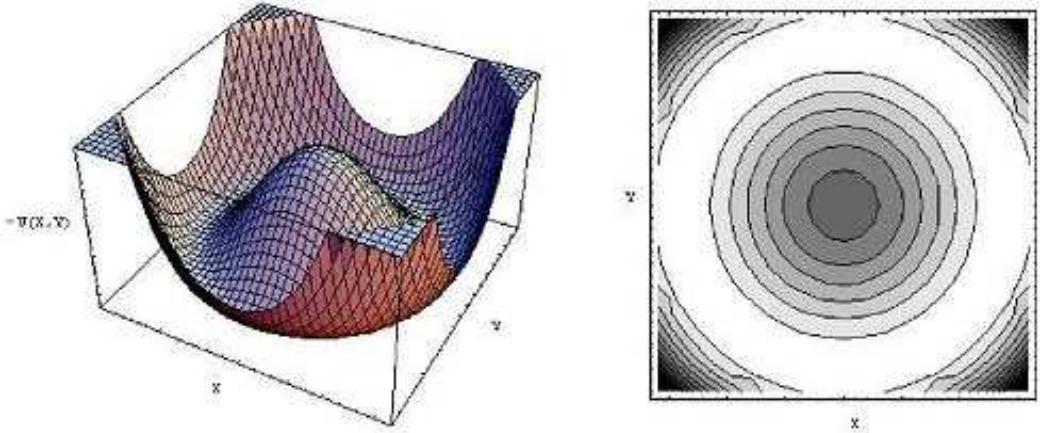}
\caption{\it Inverted potential -U(r)}
\end{center}
\end{figure}

%-------------------------------------------------------------------------------------------
\section{Analytic continuation in $\omega_c$} 
%-------------------------------------------------------------------------------------------
\indent\par 
We found useful to make analytic continuation in charge (or $\omega_c$). To present some kind of grounds to make such a procedure we cite the theorem taken from~\cite{Fedoryuk}.

{\it Theorem:}
Consider the next multiple integral:
$$
F(\lambda,\alpha)=\int_{\gamma} f(\vec{z},\alpha) \exp(\lambda S(\vec{z},\alpha)) d\vec{z}
$$
where $\alpha=(\alpha_1,...,\alpha_k)$ is a set of parameters, $\gamma$ is a contour in $\mathbf{C}^n$, with conditions that the functions $f(\vec{z},\alpha)$ and $S(\vec{z},\alpha)$ are analytic, $S(\vec{z},\alpha)$ has non-degenerate saddle points at $\vec{z}_1, ..., \vec{z}_s$ and contour $\gamma$ goes through the saddle points when $\alpha=\alpha_0$. ,  
Then the asymptotics of the integral when $\lambda\rightarrow\infty$ is given by the contribution of the saddle points $\vec{z}_1(\alpha), ..., \vec{z}_s(\alpha)$, such that $\vec{z}_1(\alpha)=\vec{z}_1, ..., \vec{z}_s(\alpha)=\vec{z}_s$, if $\alpha$ is close enough to $\alpha_0$.

It is straightforward to make generalization of the theorem cited above in the case of path integrals, implying "path integral" instead of "n-dimensional integral" and "trajectory"  instead of "n-dimensional stationary point". We have only one parameter $\omega_c$ instead of the set of parameters $\alpha=(\alpha_1,...,\alpha_k)$ and the condition $\alpha=\alpha_0$ corresponds to $\omega_c=0$. But we have a real solution of the equation of motion at $\omega_c=0$. It follows from the theorem that the asymptotics of the path integral is given by the same analytic formula inside some circle around $\omega_c=0$ on the complex plane. Thus if we calculate that one for which the stationary trajectories are real ones, the asymptotics inside all the circle can be found by continuation in $\omega_c$.

To have even more basis let us look at the convergence of the path integral. Written in the sliced form the path integral is:
\begin{multline*}
\int dx_1...dx_{N-1} \int dy_1...dy_{N-1} \exp\left\{-\epsilon \sum_{j=0}^{N-1}\left[ \frac12\left(\frac{x_{j+1}-x_j}{\epsilon}\right)^2+ \frac12\left(\frac{y_{j+1}-y_j}{\epsilon}\right)^2 + \right.\right. \\ \left.\left. + \frac{i\omega_c}{2}\left[\left(\frac{x_{j+1}-x_j}{\epsilon}\right)\left(\frac{y_{j+1}+y_j}{2}\right)-\left(\frac{y_{j+1}-y_j}{\epsilon}\right)\left(\frac{x_{j+1}+x_j}{2}\right)\right]+U(x_j,y_j) \right] \right\}= \\ =
\int dx_1...dx_{N-1} \int dy_1...dy_{N-1} \exp\left\{-\sum_{j=0}^{N-1}\left[ \frac{\left(x_{j+1}^2-2 x_{j+1}x_j+x_j^2\right)}{2\epsilon}+ \frac{\left(y_{j+1}^2-2 y_{j+1}y_j+y_j^2\right)}{2\epsilon} + \right.\right. \\ \left.\left. + \frac{i\omega_c}{4}\left[\left(x_{j+1}-x_j\right)\left(y_{j+1}+y_j\right)-\left(y_{j+1}-y_j\right)\left(x_{j+1}+x_j\right)\right]+\epsilon U(x_j,y_j) \right] \right\}
\end{multline*}
Consider the terms with the index $k$ inside the sum:
\begin{multline*}
\frac{1}{2\epsilon}\left(2 x_k^2-2 x_k x_{k-1}-2 x_{k+1}x_k\right)+ 
\frac{1}{2\epsilon}\left(2 y_k^2-2 y_k y_{k-1}-2 y_{k+1}y_k\right) + \\ + \frac{i\omega_c}{4}\left[x_{k+1}y_k - x_k y_{k+1} - x_k y_k - y_{k+1} x_k + y_k x_{k+1} +y_k x_k + \right. \\ \left. +x_k y_k +x_k y_{k-1} -x_{k-1} y_k - y_k x_k -y_k x_{k-1}+y_{k-1}x_k \right]+\epsilon U(x_k,y_k)=\\=
\frac{1}{\epsilon}\left(x_k^2-x_k (x_{k-1}+x_{k+1})-\epsilon\frac{i\omega_c}{2}x_k (y_{k+1}-y_{k-1})\right)+ \\ +
\frac{1}{\epsilon}\left(y_k^2-y_k (y_{k-1}+y_{k+1})+\epsilon\frac{i\omega_c}{2}y_k (x_{k+1}-x_{k-1})\right)+\epsilon U(x_k,y_k)
\end{multline*}
It can be seen that the term with $\omega_c$ must not affect convergence of the Gauss integrals because of small factor $\epsilon$ in front.

Thus at least for small $\omega_c$ we can reduce the stationary point (steepest-descent) method to that of Laplace, which is more simple. As a result all the equations are real and the instanton trajectories correspond to that ones of a classical particle moving in the inverted potential with a transverse magnetic field.

%-------------------------------------------------------------------------------------------
\section{Classical trajectories} 
%-------------------------------------------------------------------------------------------
\indent\par Taking into account the considerations above and making transformation to imaginary times $t\rightarrow -i\tau$ as well as $\omega_c\rightarrow i\omega_c$ (in fact the sign in the last procedure must not be the matter) we get the following Lagrangian:
\begin{equation}
L=\frac{\dot{x}^2+\dot{y}^2}{2}-\frac{\omega_c}{2}(\dot x y - \dot y x) + U(r)
\label{L}
\end{equation}

\begin{figure}[htb!]
\begin{center}
\includegraphics{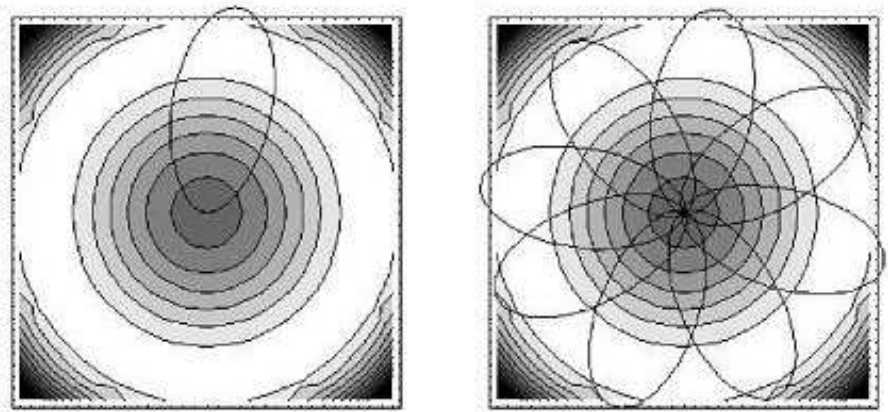}
\caption{\it The instanton trajectory slides down from the hill at the center with almost zero velocity, bounces from the wall drawing a hint and returns back to the origin in infinite time. There could be many of them, differing both in time and angular position.}
\end{center}
\end{figure}

The trajectories extremizing the action corresponding to this Lagrangian are that ones, which correspond to a classical particle moving in the inverted  potential $-U(r)$ and transverse magnetic field applied in the same direction as before. Now we are going to make a little step back from the promise to work in cartesian coordinates and find the classical trajectories from the equations of motion written in polar coordinates. Thus:
$$ L=\frac{\dot{r}^2+r^2 \dot\phi^2}{2}+\frac{\omega_c}{2}r^2\dot\phi + U(r) $$
equations of motion:
$$ r^2(\dot\phi+\frac{\omega_c}{2})=const $$
$$ \ddot{r}-r\phi^2 - \omega^2 r + 4 \alpha r^3 - \omega_c r \dot\phi=0 $$
We consider the limit $T\rightarrow\infty$ as usual for description of ground state decay in the instanton approach. The classical trajectories that give the greatest contribution to the path integral are that ones, that spend almost all their time at the origin, as the action is zero there. This gives $E=0$. Let us consider the first equation of motion and the energy conservation law:
$$r^2(\dot\phi+\frac{\omega_c}{2})=const $$
$$ 0=E=\frac{\dot{r}^2+r^2 \dot\phi^2}{2} - U(r) $$
Because the trajectory comes from the center of the system, $r\rightarrow0$ gives $const=0$. Suppose the opposite. Then at least $\dot\phi\sim 1/r^2$ when $r\rightarrow0$. Obviously this contradicts to the energy conservation law as the potential tends to zero under this limit. Hence $\dot\phi=-\omega_c/2$ and the equations of motion transform to:
$$ \ddot{r} - \Omega^2 r + 4 \alpha r^3=0 $$
with $\Omega^2=\omega^2-\omega_c^2/4$.
The solution of this equation is:
$$
r_{cl}(\tau)=\frac{\Omega}{\sqrt{2\alpha}\cosh{\Omega \tau}}
$$
It corresponds to the instanton with the center at $\tau=0$. There are many other classical trajectories with different positions of the centers. Obviously, all of them have the same action:
\begin{equation}
S_{cl}=\frac{\Omega^3}{3\alpha}
\label{Scl}
\end{equation}
We will take them into account when taking integral over time zero-mode below. Also, there are trajectories with the same position in time, but differing from each other by rotation around the origin. Similarly, these ones will be counted by the integral over $\phi$-mode below.

The instanton trajectory transformed to cartesian coordinates reads:
\begin{equation}
\begin{cases}
x_{cl}(\tau)=r_{cl}(\tau)\cos\left(-\frac{\omega_c\tau}{2}+\phi_0\right)= \frac{\Omega\cos\left(-\frac{\omega_c\tau}{2}+\phi_0\right)}{\sqrt{2\alpha}\cosh{\Omega \tau}}
\\
y_{cl}(\tau)=r_{cl}(\tau)\sin\left(-\frac{\omega_c\tau}{2}+\phi_0\right)= \frac{\Omega\sin\left(-\frac{\omega_c\tau}{2}+\phi_0\right)}{\sqrt{2\alpha}\cosh{\Omega \tau}}
\end{cases}
\label{cl_tr}
\end{equation}
Where $1/\Omega$ plays a role of the "lifetime" of the instanton, $\omega_c/2$ is the frequency of rotation around the center. For instance, when $\omega_c/2\sim\Omega$ the instanton makes approximately one turn during his "life", while for bigger $\omega_c/2$ the trajectories become spirals spinning around the center (although in this considerations $\omega_c$ is not small already, still it is worthwhile to consider such trajectories by the reasons given below).

%-------------------------------------------------------------------------------------------
\section{Jacobi fields} 
%-------------------------------------------------------------------------------------------
\indent\par Let us calculate the contribution of quantum fluctuations near the classical trajectories. Obviously, there is a set of them associated with different $\phi_0$. We can fix the one corresponding to $\phi_0=0$. The other classical trajectories, as well as the fluctuations around them, will be taken into account later integrating over the rotation symmetry group.

In the semiclassical approximation the action is decomposed about the classical trajectory (for sake of simplicity we omit the normalization constants in front of path integrals):
\begin{equation}
\int D \vec{r}(\tau) e^{-S[\vec{r}]}=e^{-S_{cl}} \int D \delta\vec{r}(\tau) e^{-\frac12 \delta^2 S}
\label{path integral}
\end{equation}
where
$$
\vec{r}(\tau)=\vec{r_{cl}}(\tau)+\delta\vec{r}(\tau)
\quad \text{and} \quad
\delta^2 S=\int_{-T}^T (\delta\vec{r},\hat{A} \delta\vec{r}) d\tau 
$$
The operator $\hat{A}$ inside the second variation of the Euclidean action is:
$$
\hat{A}=
\begin{pmatrix} 
      -\partial_\tau^2+U_x'' & \omega_c \partial_\tau + U_{xy}'' \\ 
      -\omega_c \partial_\tau + U_{xy}'' & -\partial_\tau^2+U_y''
\end{pmatrix}
$$
with
$$ U_x''=-\frac{2\Omega^2+4\Omega^2 \cos^2\frac{\omega_c \tau}{2}}{\cosh^2\Omega\tau}+\Omega^2+\omega_c^2/4 $$
$$ U_y''=-\frac{2\Omega^2+4\Omega^2 \sin^2\frac{\omega_c \tau}{2}}{\cosh^2\Omega\tau}+\Omega^2+\omega_c^2/4 $$
$$ U_{xy}''= \frac{4\Omega^2}{\cosh^2\Omega\tau} \sin\frac{\omega_c \tau}{2}\cos\frac{\omega_c \tau}{2}=\frac{2\Omega^2}{\cosh^2\Omega\tau} \sin(\omega_c \tau)$$
being the second derivatives of the potential evaluated along the classical trajectory. It was convenient here to express the frequency $\omega$ via $\Omega$ and $\omega_c$.

An arbitrary quantum deviation can be decomposed through normalized eigenfunctions $\vec\chi_i$ of the operator $\hat{A}$:
\begin{equation}
\delta\vec{r}(\tau)=\sum C_i \vec\chi_i 
\label{chi_i}
\end{equation}
Substitution to the path integral leads to the Gaussian integrations over the coefficients~$C_i$:
$$
\int D \delta\vec{r}(\tau) e^{-\frac12 \delta^2 S}=
(\sqrt{2\pi})^N (\det '\hat{A})^{-1/2} \int \frac{dC_\phi}{\sqrt{2\pi}} \int \frac{dC_\tau}{\sqrt{2\pi}} \int \frac{dC_{-}}{\sqrt{2\pi}} e^{-\frac12 C_{-}^2 \lambda_{-}}
$$
where $\det'\hat{A}$ denotes product of the eigenvalues of $\hat{A}$ omitting the zero eigenvalues $\lambda_\phi$, $\lambda_\tau$ and the negative one $\lambda_{-}$. They require special treatment and we will pay attention to them in the next sections.

It is convenient to express the resulting survival amplitude $G(\vec{0},T;\vec{0},-T)$ in terms of that one for pure parabolic well, that coincides with the contribution of the trivial classical trajectory $\vec{r_{cl}}\equiv 0$ up to the second order:
\begin{equation}
\frac{Z_1}{Z_0}=e^{-S_{cl}} \left[ \frac{\det'\hat{A}}{\det\hat{A_0}}\right]^{-1/2} \int \frac{dC_\phi}{\sqrt{2\pi}} \int \frac{dC_\tau}{\sqrt{2\pi}} \int \frac{dC_{-}}{\sqrt{2\pi}} e^{-\frac12 C_{-}^2 \lambda_{-}}
\label{z1z0}
\end{equation}
in this formula $Z_1$ defines single instanton contribution, while $Z_0$ is reserved for the trivial trajectory.

In the multidimensional case~\cite{nDim} the ratio of the determinant $\det'\hat{A}$ and $\det\hat{A_0}$ can be expressed through that one of the determinants $J$ and $J_0$ of Jacobi fields:
\begin{equation}
\frac{\det'\hat{A}}{\det\hat{A_0}}=\frac{J}{J_0\lambda_\phi \lambda_\tau \lambda_{-}}
\label{dets}
\end{equation}
The Jacobi fields satisfy:
\begin{equation}
\frac{d}{dt}\left( \frac{\partial^2 L}{\partial\dot{x}_i\partial\dot{x}_l}\dot{J}_{lk} \right)
+ \left( \frac{\partial^2 L}{\partial\dot{x}_i\partial x_l} - \frac{\partial^2 L}{\partial x_i\partial \dot{x_l}} \right)\dot{J}_{lk} + \left[ \frac{d}{dt}\left(\frac{\partial^2 L}{\partial\dot{x}_i \partial x_l} \right)-\frac{\partial^2 L}{\partial x_i \partial x_l} \right]J_{lk}=0
\label{Jacobi_fields}
\end{equation}
with boundary conditions
$$
J_{ik}=0, \;\;\; \frac{\partial J_{ik}}{\partial t}=\frac{1}{m}\delta_{ik}=\delta_{ik}
$$
The determinant $J_0$ as well as the eigenvalues in the formula~\eqref{dets} will be calculated later. Now we begin with evaluation of $J$.

The system of four differential equations of the second order ~\eqref{Jacobi_fields} written for the Lagrangian~\eqref{L} decouples into two subsystems, each one of the form:
\begin{equation}
\begin{cases} 
       -\ddot{\xi}+\omega_c \dot{\eta}+U_x'' \xi + U_{xy}''\eta=0 \\
       -\ddot{\eta}-\omega_c \dot{\xi}+U_y'' \eta + U_{xy}''\xi=0 
\end{cases} 
\label{syst}
\end{equation}
or in terms of the operator $\hat{A}$ introduced above:
$$
\hat{A}\vec{\varphi}(\tau)=0, \quad \text{with}\quad \vec{\varphi}(\tau)=\begin{pmatrix}\xi \\ \eta\end{pmatrix}
$$
The boundary conditions are:
\begin{equation}
\xi(-T)=0, \;\;\; \dot\xi(-T)=1, \;\;\; \eta(-T)=0, \;\;\; \dot\eta(-T)=0
\label{boundary1}
\end{equation}
for the first subsystem (for which $\xi\equiv J_{xx}, \eta\equiv J_{yx}$), and
\begin{equation}
\xi(-T)=0, \;\;\; \dot\xi(-T)=0, \;\;\; \eta(-T)=0, \;\;\; \dot\eta(-T)=1
\label{boundary2}
\end{equation}
for the second one ($\xi\equiv J_{yx}, \eta\equiv J_{yy}$).

Let us find 4 independent solutions of the system. Two solutions of this problem are the zero eigenmodes, corresponding to the $\tau$- and $\phi$-symmetries. They can be easily found by differentiating the classical trajectory~\eqref{cl_tr} with respect to $\tau$ and $\phi_0$. One can check by straightforward substitution that the following solutions are indeed zero-modes of the system:

$\phi$-mode:
$$ \vec{\varphi_1}(\tau)\equiv \begin{pmatrix} 
      \xi_1 \\ 
      \eta_1
    \end{pmatrix} =
    \frac{1}{\cosh\Omega \tau} 
    \begin{pmatrix} 
      \sin\frac{\omega_c \tau}{2} \\ 
      \cos\frac{\omega_c \tau}{2}
    \end{pmatrix}
$$

$\tau$-mode:
$$ \vec{\varphi_2}(\tau)\equiv \begin{pmatrix} 
      \xi_2 \\ 
      \eta_2
    \end{pmatrix} =
    \frac{\sinh\Omega\tau}{\cosh^2\Omega \tau}
    \begin{pmatrix} 
      -\cos\frac{\omega_c \tau}{2}\\ 
      \sin\frac{\omega_c \tau}{2}
    \end{pmatrix}
 $$
where we have chosen the time zero mode without the $\phi$-shifting term (obviously it does not change anything -- any two linear combinations of them could be chosen) and $\phi_0=0$.

There are two more solutions left. It could seem sophisticated to find them, however we will use the following trick. Suppose we deal with a 1-dimensional case. If one solution of the second order homogeneous differential equation (written in the Liouville form) is:
$$ f=\frac{1}{\cosh\Omega \tau} $$
then the second one can be found as (for example~\cite{Butkov} or~\cite{Kleinert}-2.7.4):
$$
g=f\int^t \frac{dt'}{f(t')^2}=\frac{\sinh\Omega\tau}{2\Omega}+\frac{\tau}{2\cosh\Omega\tau}
$$
and if
$$ f=\frac{\sinh\Omega\tau}{\cosh^2\Omega \tau} $$
the second one is
$$
g=f\int^t \frac{dt'}{f(t')^2}=\frac{1}{\cosh\Omega \tau}=\frac{\sinh^2\Omega\tau}{2\Omega\cosh\Omega\tau}+\frac{3}{2}\tau\frac{\sinh\Omega\tau}{\cosh^2\Omega\tau}-\frac{1}{\Omega\cosh\Omega\tau}
$$
The idea to take the anzats in the similar form multiplied by sines or cosines turns out to be successful. Indeed, making a straightforward substitution of functions
$$ \vec{\varphi_3}(\tau)\equiv\begin{pmatrix} 
      \xi_3 \\ 
      \eta_3
    \end{pmatrix} =
    \left[\frac{\sinh\Omega\tau}{2\Omega}+\frac{\tau}{2\cosh\Omega\tau}\right]
    \begin{pmatrix} 
      \sin\frac{\omega_c \tau}{2} \\ 
      \cos\frac{\omega_c \tau}{2}
    \end{pmatrix}
$$
$$ \vec{\varphi_4}(\tau)\equiv\begin{pmatrix} 
      \xi_4 \\ 
      \eta_4
    \end{pmatrix} = \left[\frac{\sinh^2\Omega\tau}{2\Omega\cosh\Omega\tau}+\frac{3}{2}\tau\frac{\sinh\Omega\tau}{\cosh^2\Omega\tau}-\frac{1}{\Omega\cosh\Omega\tau}\right]
    \begin{pmatrix} 
      -\cos\frac{\omega_c \tau}{2} \\ 
      \sin\frac{\omega_c \tau}{2}
    \end{pmatrix}
$$
into the system~\eqref{syst}, we conclude that they are indeed the solutions we were looking for.

Let us analyze the properties of this 4 independent solutions. It is easy to note that:
\begin{equation}
\eta_1(\tau),\; \eta_2(\tau),\; \xi_3(\tau),\; \xi_4(\tau) \text{  are EVEN functions: }
\label{eo}
\end{equation}
$$
\xi_1(\tau),\;  \xi_2(\tau),\; \eta_3(\tau),\; \eta_4(\tau)  \text{  are ODD functions: } 
$$
These important properties will be used in future for further calculations.

At last we are able to find the solution of the system, satisfying the desired boundary conditions~\eqref{boundary1} and~\eqref{boundary2}. Expanding
$$
\vec{\varphi}(\tau)=\sum_{i=1}^{4} c_i \vec{\varphi_i}(\tau)
$$
we get the coefficients
$$
c_1=-\xi_3(-T), \;\;\; c_2=-\xi_4(-T), \;\;\; c_3=\xi_1(-T), \;\;\; c_4=\xi_2(-T)
$$
for the vector $\vec{\varphi}(\tau)=\begin{pmatrix} J_{xx} \\ J_{yx}\end{pmatrix}$, and
$$
d_1=-\eta_3(-T), \;\;\; d_2=-\eta_4(-T), \;\;\; d_3=\eta_1(-T), \;\;\; d_4=\eta_2(-T)
$$
for $\vec{\varphi}(\tau)=\begin{pmatrix} J_{xy} \\ J_{yy}\end{pmatrix}$. 

Now let us look at their behavior in the limit $T\rightarrow\infty$. For simplicity we can choose $T$ in such a way that $\sin\frac{\omega_c T}{2}=0$ and $\cos\frac{\omega_c T}{2}=1$. Obviously, the final result must not depend on our specific choice of $T$, hence choosing $T$ in this way we get significant simplification:
\begin{equation}
\xi_1(T)=0, \quad  \xi_2(T)\simeq-2 e^{-\Omega T}, \quad  \xi_3(T)=0, \quad   \xi_4(T)\simeq-\frac{1}{4\Omega}e^{\Omega T},
\label{asympt+T}
\end{equation}
\begin{equation}
\quad \eta_1(T)\simeq2e^{-\Omega T}, \quad \eta_2(T)=0, \quad  \eta_3(T)\simeq\frac{1}{4\Omega}e^{\Omega T}, \quad  \eta_4(T)=0 
\label{asympt-T}
\end{equation}
and
$$
\xi_1(-T)=0, \quad  \xi_2(-T)\simeq 2 e^{-\Omega T}, \quad  \xi_3(-T)=0, \quad   \xi_4(-T)\simeq-\frac{1}{4\Omega}e^{\Omega T},$$
$$ \quad \eta_1(-T)\simeq2e^{-\Omega T}, \quad \eta_2(-T)=0, \quad  \eta_3(-T)\simeq-\frac{1}{4\Omega}e^{\Omega T}, \quad  \eta_4(-T)=0 $$
Bearing this in mind we rewrite our coefficients in the following asymptotic form:
$$
c_1=0, \;\;\; c_2\simeq\frac{1}{4\Omega}e^{\Omega T}, \;\;\; c_3=0, \;\;\; c_4\simeq2e^{-\Omega T}
$$
$$
d_1\simeq\frac{1}{4\Omega}e^{\Omega T}, \;\;\; d_2=0, \;\;\; d_3\simeq2e^{-\Omega T}, \;\;\; d_4=0
$$
Finally, the determinant $J$ can be found:
$$
J=\det
\begin{pmatrix} 
      J_{xy} & J_{xy} \\ 
      J_{yx} & J_{yy}
\end{pmatrix}
\simeq -\frac{1}{\Omega^2}
$$

%-------------------------------------------------------------------------------------------
\section{Elimination of zero eigenvalues} 
%-------------------------------------------------------------------------------------------
\indent\par Now we need to get rid of the zero eigenvalues in the determinant $J$. This could be done by several ways. One could introduce eigenvalue $\lambda$ as a small parameter, perturbating the system of differential equations and take the limit $\lambda\rightarrow 0$ of the determinator divided by $\lambda^2$ at the end (as we have two zero eigenvalues). Note, that this limit must be taken {\it after} the limit $T\rightarrow\infty$, as the zero eigenvalues are not exactly zero but tend to it as an exponential of $T$. However this requires the precision at least $o(\lambda^2)$. Thus it seems more convenient to eliminate $\lambda_\phi$ and $\lambda_\tau$ separately using the boundary perturbation method, just looking at the behavior of this eigenvalues at large $T$. As was mentioned above we must expect exponential dependence of $T$.

Consider the Green function:
\begin{equation}
\hat{A}\hat{G}(\tau,\tau')=-\hat{I}\delta(\tau-\tau')
\label{Green_eq}
\end{equation}
The reason to introduce the minus sign into the definition will be clear below: we will receive just the same boundary conditions for the system of differential equations that was solved earlier. The boundary conditions for the Green function are:
\begin{equation}
\hat{G}(-T,\tau')=\hat{0}, \;\;\; \frac{\partial\hat{G}}{\partial\tau}(-T,\tau')=\hat{0} \quad \forall \tau'
\label{Green_ini}
\end{equation}
Then the general solution of $\hat{A}\vec{\psi}=\lambda_0 \vec{\psi}$ is
$$
\vec{\psi}=\vec{\psi_0}-\lambda_0 \int_{-T}^{T} \hat{G}(\tau,\tau')\vec{\psi}(\tau')d\tau'
$$
where $\vec{\psi_0}$ is a solution of the homogeneous equation $\hat{A}\vec{\psi}_0=0$.
The eigenvalues $\lambda_\phi$ and $\lambda_t$ can be calculated requiring $\vec{\psi}(-T)$ and $\vec{\psi}(T)$ to be strictly zero (note, that the zero modes that we have found early do not satisfy this boundary conditions exactly for finite values of $T$, but only in the limit $T\rightarrow\infty$).
This leads to the following conditions:
$$
\vec{\psi_0}(-T)=0 \quad\text{and}\quad \vec{\psi_0}(T)-\lambda_0 \int_{-T}^{T} \hat{G}(T,\tau')\vec{\psi_0}(\tau')d\tau'=0
$$
where we have made the Born approximation substituting $\vec{\psi}$ by $\vec{\psi_0}$ inside the integral as $\lambda_0$ is small.

First let us find $\lambda_0\equiv\lambda_\phi$. Then the solution of homogeneous system can be expressed via $\phi$- zero mode and other solutions as:
$$
\vec{\psi_0}=\vec{\varphi_1}+\alpha \vec{\varphi_2}+\beta \vec{\varphi_3}+\gamma \vec{\varphi_4}
$$
and the boundary conditions become
\begin{equation}
\begin{cases}
\vec{\varphi_1}(-T)+\alpha \vec{\varphi_2}(-T)+\beta \vec{\varphi_3}(-T)+\gamma \vec{\varphi_4}(-T)=0
\\
\vec{\varphi_1}(T)+\alpha \vec{\varphi_2}(T)+\beta \vec{\varphi_3}(T)+\gamma \vec{\varphi_4}(T)=\lambda_\phi \int_{-T}^{T} \hat{G}(T,\tau')\vec{\varphi_1}(\tau') d\tau'
\end{cases}
\label{phi-mode-equations}
\end{equation}
In the last formula the corrections were neglected inside the integral, as they give less contribution than $\vec{\varphi_1}$ (more strictly one can find the values of the coefficients from the equations below, substitute them into this integral and prove that it is indeed the case).

Taking the same $T$, chosen so that $\sin\frac{\omega_c T}{2}=0$ and $\cos\frac{\omega_c T}{2}=1$, we get:
$$
\begin{cases}
\begin{pmatrix} 0 \\ 2 e^{-\Omega T} \end{pmatrix}
+\alpha \begin{pmatrix} 2 e^{-\Omega T} \\ 0 \end{pmatrix}
+\beta \begin{pmatrix} 0 \\ -\frac{1}{4\Omega}e^{\Omega T} \end{pmatrix}
+\gamma \begin{pmatrix} -\frac{1}{4\Omega}e^{\Omega T} \\ 0 \end{pmatrix} =0
\\
\begin{pmatrix} 0 \\ 2 e^{-\Omega T} \end{pmatrix}
+\alpha \begin{pmatrix} -2 e^{-\Omega T} \\ 0 \end{pmatrix}
+\beta \begin{pmatrix} 0 \\ \frac{1}{4\Omega}e^{\Omega T} \end{pmatrix}
+\gamma \begin{pmatrix} -\frac{1}{4\Omega}e^{\Omega T} \\ 0 \end{pmatrix}
=\lambda_\phi \int_{-T}^{T} \hat{G}(T,\tau')\vec{\varphi_1}(\tau') d\tau'
\end{cases}
$$

Note, that we have the products of two functions of $\tau'$ inside the integral, that are either even or odd~\eqref{eo}. Thus only that terms contribute, that consist of functions, both even or odd at the same time. Using the explicit form the Green function from the next section~\eqref{Greenf} we get the following expression for the integral:
$$
\int_{-T}^{T} \hat{G}(T,\tau')\vec{\varphi_1}(\tau') d\tau'=\frac{e^{\Omega T}}{4\Omega}\int_{-T}^{T} \begin{pmatrix} -\xi_2(\tau')\xi_1(\tau')-\eta_2(\tau')\eta_1(\tau') \\ \xi_1(\tau')^2+\eta_1(\tau')^2  \end{pmatrix} d\tau'
$$

The two equations from which the desired eigenvalue can be determined are:
$$
\begin{cases} 
2 e^{-\Omega T}-\beta \frac{1}{4\Omega}e^{\Omega T}=0 \\
2 e^{-\Omega T}+\beta \frac{1}{4\Omega}e^{\Omega T}=\lambda_\phi \frac{e^{\Omega T}}{4\Omega}\int_{-T}^{T} \left( \xi_1(\tau')^2+\eta_1(\tau')^2 \right) d\tau'
\end{cases}
$$
In the limit of large $T$ this leads to (see Appendix):
$$ \lambda_\phi\simeq 8 \Omega^2 e^{-2\Omega T} $$

Using the same procedure we find the second zero-eigenvalue $\lambda_\tau$. Decomposing the solution of the homogeneous system as
$$
\vec{\psi_0}=\vec{\varphi_2}+\alpha \vec{\varphi_1}+\beta \vec{\varphi_3}+\gamma \vec{\varphi_4}
$$
we write the boundary condition:
\begin{equation}
\begin{cases}
\vec{\varphi_2}(-T)+\alpha \vec{\varphi_1}(-T)+\beta \vec{\varphi_3}(-T)+\gamma \vec{\varphi_4}(-T)=0
\\\vec{\varphi_2}(T)+\alpha \vec{\varphi_1}(T)+\beta \vec{\varphi_3}(T)+\gamma \vec{\varphi_4}(T)=\lambda_\tau \int_{-T}^{T} \hat{G}(T,\tau')\vec{\varphi_2}(\tau') d\tau'
\end{cases}
\label{tau-mode-equations}
\end{equation}
Eventually, we get:
$$ \lambda_\tau \simeq 24 \Omega^2 e^{-2\Omega T} $$
Note, that $\lambda_\tau$ has the same form as for the 1-dimensional case~\cite{Schulman} (there $\epsilon\rightarrow\frac{\Omega^2}{2}$ and the operator, for which the eigenvalue is calculated is defined two times smaller than our $\hat A$), but the frequency $\Omega$ is dependent on magnetic field now: $\Omega=\sqrt{\omega^2-\omega_c^2/4}$.

%-------------------------------------------------------------------------------------------
\section{Calculation of the Green function} 
%-------------------------------------------------------------------------------------------
\indent\par
The Green function, satisfying~\eqref{Green_eq} and~\eqref{Green_ini} has the next matrix form:
$$
\hat{G}(\tau,\tau')=
\begin{pmatrix}
	g_{11} & g_{12} \\
	g_{21} & g_{22}
\end{pmatrix}
$$
where $g_{ij}=g_{ij}(\tau,\tau')$.

The equation~\eqref{Green_eq} splits into two independent subsystems:
$$
\hat{A}\begin{pmatrix} g_{11} \\ g_{21} \end{pmatrix}=\begin{pmatrix} -\delta (\tau-\tau') \\ 0 \end{pmatrix}
$$
and
$$
\hat{A}\begin{pmatrix} g_{12} \\ g_{22} \end{pmatrix}=\begin{pmatrix} 0 \\ -\delta (\tau-\tau') \end{pmatrix}
$$
with zero initial conditions following from~\eqref{Green_ini} for both of them. 

Consider the first one. It corresponds to the homogeneous system considered above in the regions $\tau<\tau'$ and $\tau>\tau'$. In that terms: $g_{11}(\tau,\tau')\equiv\xi(\tau)$ and $g_{21}(\tau,\tau')\equiv\eta(\tau)$. Both "left" and "right" solutions must be connected in such a way, that the delta function comes out. If we integrate the system over a small interval with the joint inside:
$$
\begin{cases} 
       -\dot{\xi}|_{\tau'-\epsilon}^{\tau'+\epsilon}+\omega_c \eta|_{\tau'-\epsilon}^{\tau'+\epsilon}+\int_{\tau'-\epsilon}^{\tau'+\epsilon} (U_x'' \xi + U_{xy}''\eta)=-1 \\
       -\dot{\eta}|_{\tau'-\epsilon}^{\tau'+\epsilon}-\omega_c \xi|_{\tau'-\epsilon}^{\tau'+\epsilon}+\int_{\tau'-\epsilon}^{\tau'+\epsilon}(U_y'' \eta + U_{xy}''\xi)=0 
\end{cases} 
$$
as $\xi$ and $\eta$ are continuous on the joint, the last terms at the left hand sides of the equations tend to zero. Obviously, the solution in the "left" region ($\tau<\tau'$) is the trivial one: $\xi(\tau)|_{\tau<\tau'}=0, \quad \eta(\tau)|_{\tau<\tau'}=0$. Thus we get the initial conditions for the "right" one:
$$
\begin{cases} 
      \dot{\xi}(\tau)|_{\tau=\tau'+\epsilon}=1 \\
      \dot{\eta}(\tau)|_{\tau=\tau'+\epsilon}=0
\end{cases} 
$$
Together with two other conditions coming from the continuity on the border of two regions, we have:
\begin{equation*}
\xi(\tau')=0, \;\;\; \dot\xi(\tau')=1, \;\;\; \eta(\tau')=0, \;\;\; \dot\eta(\tau')=0
\end{equation*}
With the replacement $\tau'\rightarrow -T$ this boundary conditions exactly coincide with~\eqref{boundary1} (and~\eqref{boundary2} for the second subsystem). Thus the coefficients (that actually are functions of $\tau'$):
$$
c_1=-\xi_3(\tau'), \;\;\; c_2=-\xi_4(\tau'), \;\;\; c_3=\xi_1(\tau'), \;\;\; c_4=\xi_2(\tau')
$$
$$
d_1=-\eta_3(\tau'), \;\;\; d_2=-\eta_4(\tau'), \;\;\; d_3=\eta_1(\tau'), \;\;\; d_4=\eta_2(\tau')
$$
Hence the Green function becomes
$$
\hat{G}(\tau,\tau')=
\begin{pmatrix}
	g_{11} & g_{12} \\
	g_{21} & g_{22}
\end{pmatrix}=
\theta(\tau-\tau') \left( \sum_{i=1}^{4} c_i(\tau') \vec{\varphi_i}(\tau), \sum_{i=1}^{4} d_i(\tau') \vec{\varphi_i}(\tau) \right)
$$
Using the asymptotic forms~\eqref{asympt+T} for $\tau=T$ we have:
\begin{equation}
\hat{G}(T,\tau')\simeq
\theta(T-\tau')
\begin{pmatrix}
	 -\frac{1}{4\Omega}e^{\Omega T}\xi_2(\tau')+2e^{-\Omega T}\xi_4(\tau') & -\frac{1}{4\Omega}e^{\Omega T}\eta_2(\tau')+2e^{-\Omega T}\eta_4(\tau') \\
	\frac{1}{4\Omega}e^{\Omega T}\xi_1(\tau')-2e^{-\Omega T}\xi_3(\tau') & \frac{1}{4\Omega}e^{\Omega T}\eta_1(\tau')-2e^{-\Omega T}\eta_3(\tau')
\end{pmatrix}
\label{Greenf}
\end{equation}

%-------------------------------------------------------------------------------------------
\section{Integrals over the peculiar eigenmodes} 
%-------------------------------------------------------------------------------------------
\indent\par
Having found the determinator $J$ and the zero eigenvalues, the integrals over zero modes left to be evaluated explicitly, as they require special treatment. Recall that the $\phi$- and $\tau$- modes were found differentiating the classical trajectory $\vec{r}_{cl}(\tau)$. We can express $\dot\vec{r}_{cl}$ through these modes:
$$
\dot\vec{r}_{cl}=\frac{\Omega}{\sqrt{2\alpha}}\left[-\frac{\omega_c}{2}\vec{\varphi_1} +\Omega\vec{\varphi}_2 \right]
$$
$$
\frac{\partial\vec{r}_{cl}}{\partial\phi_0}|_{\phi_0=0}=\frac{\Omega}{\sqrt{2\alpha}}\vec{\varphi_1}
$$
so that:
\begin{multline*}
\vec{r}_{cl}(\tau+d\tau)|_{\phi_0=d\phi}\simeq\vec{r}_{cl}(\tau)|_{\phi_0=0}+\dot\vec{r}_{cl}d\tau +\frac{\partial\vec{r}_{cl}}{\partial\phi_0}|_{\phi_0=0}d\phi= \\
\vec{r}_{cl}(\tau)|_{\phi_0=0}
-\frac{\Omega}{\sqrt{2\alpha}}\frac{\omega_c}{2}\vec{\varphi_1}d\tau +\frac{\Omega}{\sqrt{2\alpha}}\Omega\vec{\varphi}_2 d\tau+
\frac{\Omega}{\sqrt{2\alpha}}\vec{\varphi_1}d\phi
\end{multline*}
This Teylor series needs to be compared with our expansion~\eqref{chi_i} over the normalized eigenfunction of the operator $\hat{A}$:
$$
\vec{r}=\vec{r_{cl}}+ C_\phi \vec\chi_\phi+ C_\tau \vec\chi_\tau+...=
\vec{r_{cl}}+ C_\phi \frac{\vec\varphi_1}{||\vec\varphi_1||}+ C_\tau \frac{\vec\varphi_2}{||\vec\varphi_2||}+...
$$
where the zero eigenmodes were extracted from the sum explicitly. In order to complete the integration over the zero modes, the integrals over the coefficients $C_\phi$, $C_\tau$ must be transformed to the integrals over angular and time positions. The expressions above make clear the procedure of changing variables from $C_\phi$, $C_\tau$ to $\phi$, $\tau$:
$$
\begin{cases}
\frac{dC_\varphi}{||\vec\varphi_1||}=
-\frac{\Omega}{\sqrt{2\alpha}}\frac{\omega_c}{2}d\tau 
+\frac{\Omega}{\sqrt{2\alpha}}d\phi
\\
\frac{dC_\tau}{||\vec\varphi_2||}=\frac{\Omega^2}{\sqrt{2\alpha}} d\tau
\end{cases}
$$
The Jacobian of this transformation is:
$$
J_{\phi \tau}=
\det \begin{pmatrix}
	\frac{\partial C_\phi}{\partial\phi} & \frac{\partial C_\phi}{\partial\tau} \\
	\frac{\partial C_\tau}{\partial\phi} & \frac{\partial C_\tau}{\partial\tau}
\end{pmatrix}=
\det \begin{pmatrix}
	\frac{\Omega}{\sqrt{2\alpha}}||\vec\varphi_1|| & -\frac{\Omega}{\sqrt{2\alpha}}\frac{\omega_c}{2}||\vec\varphi_1|| \\
	0 &  \frac{\Omega^2}{\sqrt{2\alpha}} ||\vec\varphi_2||
\end{pmatrix}=\frac{\Omega^3}{2\alpha} ||\vec\varphi_1|| ||\vec\varphi_2||=\frac{\Omega^2}{\sqrt{3}\alpha}
$$
where we have have substituted $||\vec\varphi_1||$ and $||\vec\varphi_2||$, found to be (see Appendix):
$$
||\vec\varphi_1||=\sqrt{\frac{2}{\Omega}}
\quad \text{and} \quad
||\vec\varphi_2||=\sqrt{\frac{2}{3\Omega}}
$$

Let us pay attention to the negative eigenvalue. Physical intuition tells that there must be only one negative eigenmode, corresponding to the direction in the functional space associated with escape of the particle. One can use also oscillation theorems to determine how many eigenvalues less than $\lambda_\tau$ and $\lambda_\phi$ left. For one-dimensional Shroedinger equation this is nothing but number of nodes of the eigenfunction that tells its position number in the ordered series of corresponding eigenvalues $\lambda_0<\lambda_1<\lambda_2<...$. We have managed with the negative eigenvalue just as it was done in one-dimensional case (see~\cite{Coleman} for details). There an analytical continuation of the Gaussian integral over negative eigenmode has been used. A special feature is the factor $\frac12$, arising from a half of the Gaussian peak. The integration over the other half axis turns out to be fake. Thus the contribution of the negative eigenmode integral is:
$$
\int \frac{dC_{-}}{\sqrt{2\pi}} e^{-\frac12 C_{-}^2 \lambda_{-}}=\pm \frac{i}{2\sqrt{-\lambda_{-}}}
$$
The sign here depends on how the analytical continuation is done. In future we will omit this sign as it is not important. The precise value of $\lambda_{-}$ is not important also as it cancel with $\lambda_{-}$ in the formula for the ratio of the determinants~\eqref{dets}.

Eventually,
$$
\int \frac{dC_\phi}{\sqrt{2\pi}} \int \frac{dC_\tau}{\sqrt{2\pi}} \int \frac{dC_{-}}{\sqrt{2\pi}} e^{-\frac12 C_{-}^2 \lambda_{-}} = \frac{i}{2\sqrt{-\lambda_{-}}} \frac{1}{2\pi} \frac{\Omega^2}{\sqrt{3}\alpha} \int_0^{2\pi} d\phi \int_{-T}^{T} d\tau
$$
We will retain this integrals unsolved for some time. Note, that the integration over the angular variable $\phi$ appears {\it naturally} from the integration over $C_\phi$ along the direction in the functional space corresponding to the $\phi$-symmetry zero mode. This is nothing to do with path integral written in polar coordinates, for which an additional care must be taken.

%-------------------------------------------------------------------------------------------
\section{Contribution of the trivial trajectory} 
%-------------------------------------------------------------------------------------------
The determinant $J_0$ of the Jacobi fields along the trivial trajectory $\vec{r}_{cl}=0$ is estimated by analogy with $J$. The calculation is rather simple now.
$$
\hat{A_0}=
\begin{pmatrix} 
      -\partial_\tau^2+\Omega^2+\omega_c^2/4  & \omega_c \partial_\tau \\ 
      -\omega_c \partial_\tau & -\partial_\tau^2+\Omega^2+\omega_c^2/4 
\end{pmatrix}
$$
So that the system $\hat{A_0}\vec\varphi(\tau)=0$ reads:
$$
\begin{cases} 
       -\ddot{\xi}+\omega_c \dot{\eta}+(\Omega^2+\omega_c^2/4)  \xi =0 \\
       -\ddot{\eta}-\omega_c \dot{\xi}+(\Omega^2+\omega_c^2/4)  \eta=0 
\end{cases} 
$$
with the same boundary conditions~\eqref{boundary1} and~\eqref{boundary2}. The system can be solved by standard methods. We only write the final answer for the determinants of the Jacobi fields in case of the pure parabolic potential:
$$
J_0\simeq \frac{e^{4\Omega T}}{4\Omega^2}
$$

%-------------------------------------------------------------------------------------------
\section{Instanton gas} 
%-------------------------------------------------------------------------------------------
\indent\par
Finally, substituting the determinants and eigenvalues found above to~\eqref{z1z0} and ~\eqref{dets}, we obtain the contribution of one instanton trajectory (more strictly of a group of equivalent instantons with different time and angular positions, as we have integrated over the zero modes already):
$$
\frac{Z_1}{Z_0}=e^{-S_{cl}} \left[ \frac{J}{J_0\lambda_\phi \lambda_\tau \lambda_{-}} \right]^{-1/2} 
\frac{i}{2\sqrt{-\lambda_{-}}} \frac{1}{2\pi} \frac{\Omega^2}{\sqrt{3}\alpha} \int_0^{2\pi} d\phi \int_{-T}^{T} d\tau=
e^{-S_{cl}} K \int_0^{2\pi} d\phi \int_{-T}^{T} d\tau
$$
with 
$$
K=\frac{i2\Omega^4}{\alpha 2\pi}
$$
We must also take into account the contribution of the multi-instanton trajectories. Also, we must integrate over their positions both in time and angular space. Thus
$$
\frac{Z_n}{Z_0}=e^{-n S_{cl}} K^n \int_{0}^{2\pi}d\phi_1 \int_{-T}^{T}d\tau_1 \int_{0}^{2\pi}d\phi_2 \int_{-T}^{\tau_1}d\tau_2 \; ...\; \int_{0}^{2\pi}d\phi_n \int_{-T}^{\tau_{n-1}}d\tau_n =e^{-n S_{cl}} K^n \frac{(2\pi 2T)^n}{n!}
$$
represents the contribution of a trajectory with $n$ instantons distributed in the time interval $(-T,T)$ and arbitrary directed with respect to the angular coordinate.
Finally, we must sum all the contributions to get the survival amplitude:
$$
G(\vec{0},T;\vec{0},-T)=Z_0+Z_1+Z_2+...=Z_0\sum_{n=0}^{\infty} e^{-n S_{cl}} K^n \frac{(2\pi 2T)^n}{n!}=Z_0\exp\left(2\pi 2T K e^{-S_{cl}} \right)
$$
Thus the probability decay rate is:
\begin{equation}
\Gamma=2 \mathit{Im}E=4\pi K e^{-S_{cl}}=\frac{4\Omega^4}{\alpha}e^{-S_{cl}}
\label{Gamma}
\end{equation}
This simple expression is just that one, that we were looking for. To complete the calculation we must transform back to the real cyclotron frequencies $\omega_c\rightarrow -i\omega_c$, so that the frequency $\Omega$ becomes a square root of the sum now: $\Omega=\sqrt{\omega^2+\omega_c^2/4}$. 

We can rewrite the formula~\eqref{Gamma} in the following way using the expression for the classical action~\eqref{Scl}:
$$
\Gamma=12\Omega S_{cl} e^{-S_{cl}}=12\Omega (\sqrt{S_{cl}})^2 e^{-S_{cl}}
$$
The factor $\sqrt S_{cl}$ has appeared twice here, just as much as the number of zero modes. This is a common situation in the instanton technique (e.g. Coleman, "Aspects of Symmetry"~\cite{Coleman} p.337 gets 4 factors $\sqrt S_{cl}$ in case of four zero modes).
%-------------------------------------------------------------------------------------------
%-------------------------------------------------------------------------------------------
\section{WKB} 
%-------------------------------------------------------------------------------------------
%-------------------------------------------------------------------------------------------
\indent\par 
Now we will get the same result using different method. Even the way, in what we deal with the magnetic field is different here. The Hamiltonian corresponding to our initial system~\eqref{2Dmodel} is:
$$
\hat H=\frac{(p_x-\frac{e}{c}A_x)^2}{2}+\frac{(p_y-\frac{e}{c}A_y)^2}{2}+U(r),
\quad
U(r)=\frac{\omega^2}{2}r^2-\alpha r^4
$$
Using the same gauge $\vec{A}=(-\frac{y B}{2},\frac{x B}{2},0)$, it can be rewritten as follows:
$$
\hat H=\frac{p_x^2+p_y^2}{2}+  \frac{\omega_c^2}{8}(x^2+y^2)-\frac{\omega_c}{2}L+U(r)=\hat H_0-\frac{\omega_c}{2}\hat L
$$
where $L$ is angular momentum operator: $L=x p_y-y p_x$ and we have defined 
$$
\hat H_0=\frac{p_x^2+p_y^2}{2}+ \frac{\omega_c^2}{8}(x^2+y^2)+U(r)
$$
The important point here is that $\hat H_0$ introduced above commutes with the angular momentum: $[\hat H_0,\hat L]=0$. This is because of the rotational invariance of the system, as the potential (including the magnetic field term) in the Hamiltonian $\hat H_0$ is a pure function of radial distance $r=\sqrt{x^2+y^2}$. 

Consider the matrix $R(-\phi)$, rotating the vectors in the counterclockwise direction:
$$
R(-\phi)=\begin{pmatrix} \cos\phi & \sin\phi \\ -\sin\phi & \cos\phi \end{pmatrix}
$$
Then the angular momentum operator $\hat L$ generates rotations of the wave function:
$$
e^{-i\hat L\phi}\psi(\vec{r})=\psi(R(-\phi)\vec{r})
$$
This can be seen as follows (e.g.~\cite{VQM} p.219). One can find generator of the rotation unitary group by differentiating with respect to $\phi$:
\begin{multline*}
i\frac{d}{d\phi}\psi(R(-\phi)\vec{r})|_{\phi=0}=i\nabla\phi(R(-\phi)\vec{r})\frac{d}{d\phi}R(-\phi)\vec{r}|_{\phi=0}= \\ =
i\nabla\psi(\vec{r})\begin{pmatrix}y \\ -x\end{pmatrix}=-i\left( x\frac{\partial}{\partial y}-y\frac{\partial}{\partial x}\right)\psi(\vec{r})=\hat{L}\psi(\vec{r})
\end{multline*}
Thus $\hat L$ is indeed the generator of the unitary group of rotations.

With all this in mind, the amplitude~\eqref{G} becomes:
$$
G(\vec{0},T;\vec{0},-T)=<\vec{0}|e^{-i 2 \hat H T} |\vec{0}> =<\vec{0}|e^{-i 2 \hat H_0 T} e^{i \omega_c \hat L T} |\vec{0}>=<\vec{0}|e^{-i 2 \hat H_0 T}|\vec{0}>
$$
as the rotation of the state $|\vec{0}>$ is equal to itself. Thus the problem has reduced to finding the decay rate of the particle in the potential of the same shape with the renormalized frequency $\omega\rightarrow\sqrt{\omega^2+\omega_c^2/4}\equiv\Omega$ in absence of the magnetic field:
$$
\hat H=\frac{p_x^2+p_y^2}{2}+U_{eff}(r)
\quad \text{with} \quad 
U_{eff}(r)=\frac{\Omega^2}{2}r^2-\alpha r^4
$$
Shroedinger equation in polar coordinates reads:
$$
\left[ -\frac{\hbar^2}{2}\left(\frac{\partial^2}{\partial r^2}+\frac{1}{r}\frac{\partial}{\partial r}+\frac{1}{r^2}\frac{\partial^2}{\partial\phi^2} \right)+
U_{eff}(r)\right] \psi=E \psi
$$
with $\hat L=-i\hbar \frac{\partial}{\partial\phi}$ and $\phi=e^{i l\phi}f(r)$:
$$
\left[ -\frac{\hbar^2}{2}\left(\frac{\partial^2}{\partial r^2}+\frac{1}{r}\frac{\partial}{\partial r}-\frac{l^2}{r^2} \right)+
U_{eff}(r)\right] f(r)=E f(r)
$$
As we are interested in the tunneling from the lowest state we take $l=0$. Consider $\psi$ in the form $f(r)=e^{\frac{i}{\hbar}S(r)}$ with $S(r)=S_0-i\hbar S_1+...$ according to the standard WKB expansion:
\begin{subequations}\label{rate equations} 
    \begin{align*} 
&f(r)=e^{\frac{i}{\hbar}S(r)} \\
&f'(r)=(\frac{i}{\hbar}S_0'+S_1'+...)e^{\frac{i}{\hbar}S(r)} \\
&f''(r)=(-\frac{1}{\hbar^2}S_0'^2+\frac{2i}{\hbar}S_0' S_1'+\frac{i}{\hbar}S_0'' +...)e^{\frac{i}{\hbar}S(r)}
    \end{align*} 
\end{subequations}
Substituting to the Shroedinger equation above and equalizing terms near the same power of $\hbar$ we get:
$$
\frac{1}{2}S_0'^2+U_{eff}(r)=E
$$
$$
2 S_0' S_1' + S_0'' + \frac{1}{r}S_0'=0
$$
From now we can assume $\hbar=1$. The first equation gives:
$$
S_0=i\int\sqrt{2(U_{eff}(r)-E)}\equiv\int p(r), \quad p(r)=\sqrt{2(U_{eff}(r)-E)}
$$
Dividing the second one by $S_0'$:
$$
2 S_1' + (\ln S_0')' + \frac{1}{r}=0
$$
that is also easily integrable. Eventually, we get the underbarrier WKB wave function in polar coordinates:
$$
\psi(r)=C\frac{1}{\sqrt{r p(r)}} \exp\left(-\int_{r_1}^r p(r) dr\right)
$$
where $r_1$ is the first classical turning point corresponding to the energy $E$.
The constant $C$ can be found comparing the result with the normalized ground state wave function for the two-dimensional parabolic well:
$$
\psi(r)=\sqrt{\frac{\Omega}{\pi}}e^{-\frac{\Omega r^2}{2}}
$$
For the underbarrier WKB wave function near the bottom (but far enough from the turning point, so that WKB works already) the integral under the exponent is approximated by
$$
\int_{r_1}^{r} p(r) dr\approx\int_{r_1}^{r}\sqrt{\Omega^2 r^2- 2E}dr=\frac{\Omega r^2}{2}-\frac{E}{2\Omega}-\frac{E}{2\Omega}\ln{\frac{2r^2\Omega ^2}{E}}
$$
for $E$ as a small parameter. Taking $E=\Omega$ as the lowest energy level for two-dimensional parabolic well and equalizing the both wavefunctions, we get:
$$
C\frac{1}{\sqrt{\Omega r^2}}e^{-\left[\frac{\Omega r^2}{2}-\frac12-\frac12 \ln{2r^2\Omega }\right]}=\sqrt{\frac{\Omega}{\pi}}e^{-\frac{\Omega r^2}{2}}
$$
where $p(r)$ was also decomposed up to the lowest order: $p(r)\approx\Omega r$. All this leads to
$$
C=\sqrt{\frac{\Omega}{2\pi e}}
$$
The outgoing wave function ($r>r_2$, with $r_2$ as a second turning point outside the well):
$$
\psi(r)=C\frac{1}{\sqrt{r p(r)}} \exp\left(-\int_{r_1}^{r_2} p(r) dr+i\int_{r_2}^{r} p(r) dr\right)
$$
This is just the wave function, we are specially interested in. Using it, we can estimate density of the probability current:
\begin{multline*}
j=\frac{i}{2}(\psi \frac{\partial}{\partial r}\psi^* - \psi^* \frac{\partial}{\partial r}\psi)=\frac{i}{2}\frac{C^2}{r p(r)}\exp\left(-2\int_{r_1}^{r_2}p(r)dr\right) \left(- i p(r)- i p(r)\right) \\
=\frac{C^2}{r}\exp\left(-2\int_{r_1}^{r_2}p(r)dr\right)=\frac{\Omega}{2\pi r e }\exp\left(-2\int_{r_1}^{r_2}p(r)dr\right)
\end{multline*}
We must integrate the current density over the circumference to get the total probability current coming outside the well:
$$
D=\int_0^{2\pi}2\pi r j d\phi=\frac{\Omega}{e}e^{-W(E)}
$$
As the wave function inside the well is normalized, this expression corresponds to the transition probability through the barrier, i.e. decay rate. We have denoted $W(E)=2\int_{r_1}^{r_2} p(r) dr$. To link this result with the one, obtained by the instanton technique, we must represent the decay rate in terms of $W(0)$. This can be done as follows. We separate the integral into three parts:
$$W(E)/2=\int_{r_1}^{r_1'} p(r) dr + \int_{r_1'}^{r_2'} p(r) dr+\int_{r_2'}^{r_2} p(r) dr$$
Where the points $r_1'$ and $r_2'$ are chosen between the turning points $r_1$ and $r_2$ so, that the first integral is calculated keeping the quadratic term only of the potential, the second one is calculated making expansion in $E/U(r)$ and the last integral is calculated approximating the potential by a linear function near the turning point~$r_2$.
$$
\int_{r_1}^{r_1'} p(r) dr\approx\int_{r_1}^{r}\sqrt{\Omega^2 r^2- 2E}dr=\int_{0}^{r_1'}\sqrt{2U_{eff}(r)}-\frac{E}{2\Omega}-\frac{E}{2\Omega}\ln{\frac{2r^2\Omega^2}{E}}
$$
\begin{multline*}
\int_{r_1'}^{r_2'} p(r) dr\approx\int_{r_1'}^{r_2'}\sqrt{2U_{eff}(r)}-\int_{r_1'}^{r_2'}\frac{E}{\sqrt{2U_{eff}(r)}}dr= \\ =\int_{r_1'}^{r_2'}\sqrt{2U_{eff}(r)} + \frac{E}{\Omega}\ln\left|\frac{(\Omega+\sqrt{\Omega^2-2\alpha r_2'^2}) r_1'}{(\Omega+\sqrt{\Omega^2-2\alpha r_1'^2}) r_2'}\right|
\end{multline*}
Near the second turning point $r_2$ we approximate potential by a linear function with zero at the the point $r_0=\Omega/\sqrt{2\alpha}$, such that $U_{eff}(r_0)=0$:
$$
U_{eff}(r)\approx-\frac{\Omega^3}{\sqrt{2\alpha}}(r-r_0)
$$
so that the third integral becomes
$$
\int_{r_2'}^{r_2} p(r) dr\approx\int_{r_2'}^{r_2} \sqrt{2\left(\frac{\Omega^3}{\sqrt{2\alpha}}(r_0-r)-E\right)} dr=\int_{r_2'}^{r_0} \sqrt{2U_{eff}(r)} dr-\frac{2\alpha E}{\Omega^3}\sqrt{2(\frac{\Omega^3}{\sqrt{2\alpha}}(r_0-r_2')}
$$
Combining this terms altogether and taking the limits $r_1'\rightarrow 0$ and $r_2'\rightarrow r_0$ we obtain:
$$
W(E)\approx W(0)-\frac{E}{\Omega}-\frac{E}{\Omega}\ln\left(\frac{4 \Omega^4}{\alpha E}\right)
$$
that gives
$$
D=\frac{4\Omega^4}{\alpha} e^{-W(0)}
$$
where 
$$
W(0)=2\int_0^{r_0}\sqrt{U_{eff}(r)}dr=2\int_0^{r_0}\sqrt{\Omega^2 r^2-2\alpha r^4}dr=\int_0^{r_0^2}\sqrt{\Omega^2-2\alpha r^2}dr^2=\frac{\Omega^3}{3\alpha}\equiv S_{cl}
$$
what coincides with $S_{cl}$~\eqref{Scl}. Evidently, using the different method we have got the same result~\eqref{Gamma} that was derived by the instanton technique. 

%-------------------------------------------------------------------------------------------
\section{Discussion} 
%-------------------------------------------------------------------------------------------
\indent\par 
In deriving the formula for the decay rate using the instanton approach we have restricted ourselves to the small values of $\omega_c$. As the standard WKB results gives the same formula, we conclude, that the obtained result is more general. Therefore, it is worthwhile to look at the instanton trajectories even for $\omega_c/2 \sim \omega$. Obviously in this regime the trajectories are spirals, spinning out and in near the top of the hill. The question is how do the trajectories near the top "feel" the behavior of the potential on large distances and reproduce the right dependence of the decay rate on the 4th power term of the double well? Let us analyze this situation more carefully. During almost all the period of the instanton existence the spirals are perpendicular to the gradient direction whereas the velocities are small at the top, because of zero energy at the top. Thus both the trajectories and the action must be strongly dependent on high order derivatives of the potential, i.e. the 4th power term.

  Although this problem can be reduced to itself under zero magnetic field as was shown above, we believe that the calculations aboce could serve at least as an instructive example of how the instanton technique works in case of a two-dimensional system with transverse magnetic field.
  
  Thus, (i) The instanton technique can be successfully applied to 2D models in magnetic field. (ii) Coincidence with standard WKB was obtained. (iii) Analytical continuation in $\omega_c$ works.

%-------------------------------------------------------------------------------------------
%-------------------------------------------------------------------------------------------
\section{Vortex expulsion problem}
%-------------------------------------------------------------------------------------------
%-------------------------------------------------------------------------------------------
\indent\par Now we are able to calculate the expulsion magnetic field, at what the vortex leaves superconducting dot. Using the London approximation and the method of images, the potential well inside the disk can be estimated~\cite{Buzdin}:
$$
V(r)=\frac{d \Phi_0^2}{16 \pi^2 \lambda^2} \left[ \frac{h^2}{4}+\ln(R/\xi)-h(1-r^2)+\ln(1-r^2) \right]
$$
where $h=H\pi R^2/ \Phi_0$ and $r$ is the distance from the center in units of the radius $R$.
The minimum exists until the field $h=1$. However, due to quantum tunneling the vortex leaves the well at higher $h>1$. We approximate this potential by the inverted double well and use the formula for the decay rate calculated above to estimate the expulsion field. This leads to the next dependance on the radius of the disk on the figure. For comparision we also present the typical dependance due to thermal activation process for some fixed temperature.

\begin{figure}[htb!]
\begin{center}
\includegraphics{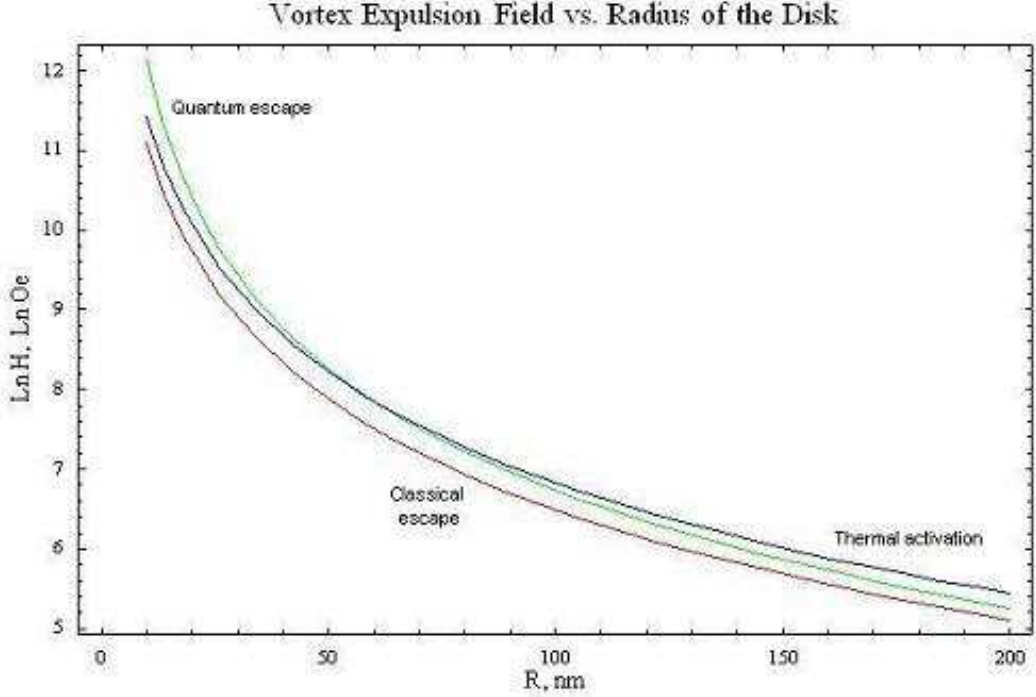}
\end{center}
\caption{\it Dependance of the expulsion field on the radius of the HTSC disk}
\end{figure}
\bigskip

\newpage
%-------------------------------------------------------------------------------------------
\section{Appendix A: Useful integrals}
%-------------------------------------------------------------------------------------------
$$
\int_{-\infty}^\infty \frac{dx}{\cosh^2 x} =2
$$

$$
\int_{-\infty}^\infty \frac{\sinh^2 x}{\cosh^4 x} dx =\frac{2}{3}
$$

$$
\int_{-\infty}^\infty \frac{1}{\cosh^4 x} dx =\frac{4}{3}
$$

%-------------------------------------------------------------------------------------------
\section{Appendix B: Faddeev-Popov procedure}
%-------------------------------------------------------------------------------------------
\indent\par 
Zero eigenvalues can be eliminated using the Faddeev-Popov procedure~\cite{Faddeev-Popov}, that is very useful for systems with constraints in quantum field theory. As our initial path integral~\eqref{path integral} is not properly defined because of overcounting of zero eigenmodes, we need to define correct measure of integration in order not to take into account the contributions from the equivalent trajectories related by symmetry group transformations (in the field theory this usually corresponds to gauge transformations). Thus, there are two constraints on our path integral:
$$ F_1[\vec{r}(\tau+\eta,\phi_0+\theta)]\equiv <\vec{r}(\tau+\eta,\phi_0+\theta),\vec\varphi_1(\tau,\phi_0)>$$
$$ F_2[\vec{r}(\tau+\eta,\phi_0+\theta)]\equiv <\vec{r}(\tau+\eta,\phi_0+\theta),\vec\varphi_2(\tau,\phi_0)>$$
Consider the identity:
\begin{equation}
\Delta[F_1[\vec{r}(\tau,\phi_0)],F_2[\vec{r}(\tau,\phi_0)]] \int_{-T}^{T} d\eta \int_0^{2\pi} d\theta \delta(F_1[\vec{r}(\tau+\eta,\phi_0+\theta)]) \delta(F_2[\vec{r}(\tau+\eta,\phi_0+\theta)]) \equiv 1 
\label{Delta}
\end{equation}
where the Faddeev-Popov determinant $\Delta[F_1,F_2]$ is the Jacobian of transformation of integration variables. $\Delta$, as well as the action along the trajectories close to the classical ones, is invariant under translations in time and $\phi$:
$$
\Delta[F_1[\vec{r}(\tau+\eta,\phi_0+\theta)],F_2[\vec{r}(\tau+\eta,\phi_0+\theta)]]=
\Delta[F_1[\vec{r}(\tau,\phi_0)],F_2[\vec{r}(\tau,\phi_0)]]
$$
Indeed, this can be easily proved by shifting variables inside the integral~\eqref{Delta}.

Introducing the identity~\eqref{Delta} to the path integral~\eqref{path integral}:
\begin{multline*}
\int D \vec{r} e^{-S[\vec{r}]}=
\\
=\int_{-T}^{T} d\eta \int_0^{2\pi} d\theta \int D \vec{r} e^{-S[\vec{r}(\tau,\phi_0)]} \Delta[F_1,F_2] \delta(F_1[\vec{r}(\tau+\eta,\phi_0+\theta)]) \delta(F_2[\vec{r}(\tau+\eta,\phi_0+\theta)])= 
\\ 
=\int_{-T}^{T} d\eta \int_0^{2\pi} d\theta \int D \vec{r} e^{-S[\vec{r}(\tau,\phi_0)]} \Delta[F_1,F_2] \delta(F_1[\vec{r}(\tau,\phi_0)]) \delta(F_2[\vec{r}(\tau,\phi_0)]) 
\end{multline*}
where we have used the invariance both of the Faddeev-Popov determinant and the action as well as the identity $D \vec{r}(\tau+\eta,\phi_0+\theta) = D \vec{r}(\tau,\phi_0)$. Thus we have have transformed our integrations over zero modes to the integration over the groups of translational symmetry.

The Faddeev-Popov determinant is found to be
\begin{multline*}
\Delta[F_1,F_2]=
\det \begin{pmatrix}
	\frac{\partial F_1}{\partial\theta} & \frac{\partial F_1}{\partial\eta} \\
	\frac{\partial F_2}{\partial\theta} & \frac{\partial F_2}{\partial\eta}
\end{pmatrix}_{\theta=0,\eta=0} =
\\
= \det \begin{pmatrix}
	<\frac{\partial \vec{r}(\tau,\phi_0)}{\partial \phi_0},\vec\varphi_1(\tau,\phi_0)>  & <\frac{\partial \vec{r}(\tau,\phi_0)}{\partial \tau},\vec\varphi_1(\tau,\phi_0)> \\
	<\frac{\partial \vec{r}(\tau,\phi_0)}{\partial \phi_0},\vec\varphi_2(\tau,\phi_0)> & <\frac{\partial \vec{r}(\tau,\phi_0)}{\partial \tau},\vec\varphi_2(\tau,\phi_0)>
\end{pmatrix}=\frac{\Omega^2}{\sqrt{3}\alpha}
\end{multline*}

%-----------------------------------------------------------------------------------------------

\end{document}